\documentclass[11pt]{article}
\usepackage{moriond,epsfig}

\def\be{\begin{equation}}
\def\ee{\end{equation}}
\def\bea{\begin{eqnarray}}
\def\eea{\end{eqnarray}}

\begin{document}
\vspace*{4cm}
\title{No Evidence for Dark Energy Dynamics from \\ a Global Analysis of Cosmological Data}

\author{P. SERRA}

\address{Center for Cosmology, Department of Physics and Astronomy,\\
University of California, Irvine, CA 92697}

\maketitle\abstracts{We use a variant of principal component analysis to investigate the possible
temporal evolution of the dark energy equation of state, $w(z)$. We constrain
$w(z)$ in multiple redshift bins, utilizing the most recent data from Type Ia
supernovae, the cosmic microwave background, baryon acoustic oscillations, the
integrated Sachs-Wolfe effect, galaxy clustering, and weak lensing data. Unlike
other recent analyses, we find no significant evidence for evolving dark energy;
the data remains completely consistent with a cosmological constant. We also
study the extent to which the time-evolution of the equation of state would be
constrained by a combination of current and future-generation surveys, such as
Planck and the Joint Dark Energy Mission.}

\section{Introduction}
One of the defining challenges for modern cosmology is 
understanding the physical mechanism responsible for the accelerating expansion
of the Universe~\cite{Riess:1998cb,Perlmutter:1998np}. The origin of the
cosmic acceleration can be due to a new source of stress-energy, called ``dark energy'', a
modified theory of gravity, or some mixture of both \cite{Uzan:2006mf}.\\
In the absence of a well-defined and theoretically motivated model for dark
energy, it is generally assumed that the dark energy equation of state (the ratio of
pressure to energy density) evolves with redshift with an arbitrary functional
form. Common parameterizations include a linear variation,
$w(z)=w_0+w_zz$~\cite{Cooray1999}, or an evolution that asymptotes to a constant
$w$ at high redshift, $w(z)=w_0+w_az/(1+z)$~\cite{Chevallier2001,Linder2003}. However, given our
complete ignorance of the underlying physical processes, it is advisable to
approach our analysis of dark energy with a minimum of assumptions. Fixing an ad
hoc two parameter form could lead to bias in our inference of the dark energy
properties.\\
In this paper we measure the evolution history of the dark energy using a
flexible and almost completely model independent 
approach, based on a variant of the principal component analysis (PCA)
introduced in Huterer (2003)~\cite{Huterer2003}; 
in order to be conservative, we begin by using data we determine the equation 
only from geometric probes of dark energy,
namely the cosmic microwave background radiation (CMB), Type Ia supernovae (SNe)
and baryon acoustic oscillation data (BAO). We perform a full likelihood analysis
using the Markov Chain Monte Carlo approach~\cite{Lewis2002}. We then consider
constraints on $w(z)$ from a larger combination of datasets, including probes of
the growth of cosmological perturbations, such as large scale structure
(LSS) data. An important consideration for such an analysis is to properly take
into account dark energy perturbations, and we make use of the prescription
introduced in~\cite{Feng2008}. This method
implements a Parameterized Post-Friedmann (PPF) prescription for the dark energy
perturbations following~\cite{Hu2007}.
 
\section{Analysis and results}\label{Analysis}
The method we use to constrain the dark energy evolution is based on a modified
version of the publicly available Markov Chain Monte Carlo package CosmoMC
\cite{Lewis2002}, with a convergence diagnostics based on the Gelman-Rubin
criterion~\cite{gelman}. We consider a flat cosmological model described by the
following set of parameters:
\begin{equation}
 \label{parameter}
      \{w_i,\omega_{b}, \omega_{c},
      \Theta_{s}, \tau,  n_{s}, \log[10^{10}A_{s}] \}~,
\end{equation}
 where $\omega_{b}$ ($\equiv\Omega_{b}h^{2}$) and $\omega_{c}$ ($\equiv\Omega_{c}h^{2}$) are the physical baryon and cold dark matter densities relative to the critical density, $\Theta_{s}$ is the ratio of the sound horizon to the angular
diameter distance at decoupling, $\tau$ is the optical depth to re-ionization,
and $A_{s}$ and $n_s$ are the amplitude of the primordial spectrum and the
spectral index, respectively. We bin the dark energy equation of state in five redshift
bins, $w_i(z)\,(i=1,2,..5)$, representing the value at five redshifts, 
$z_i\in{[0.0,0.25,0.50,0.75,1.0]}$ and, for $z>1$, we fix the equation of state parameter 
at its $z=1$ value, since we find that current data place only weak
constraints on $w(z)$ for $z>1$.  To summarize, our parameterization is given by:
\begin{equation} \label{paraw} w(z)= \left\{
    \begin{array}{ll}
          w(z=1),&  \hbox{$z> 1$;} \\
      \hbox{  $w_i$}, & \hbox{$z\leq z_{max}, z\in \{z_i\}$;} \\
      \hbox{ spline}, & \hbox{$z\leq z_{max}, z\notin \{z_i\}$.}
    \end{array}
\right. \end{equation}
Finally, we follow~\cite{HutererCooray2005} to determine
uncorrelated estimates of the dark energy parameters.\\
For the CMB, we use data and likelihood
code from the WMAP team's 5-year release~\cite{Dunkley:2008ie} (both temperature
TT and polarization TE; we will refer to this analysis as WMAP5). We also checked that results don't change if 
we use the latest data release from WMAP \cite{Komatsu2010}.
Supernova data come from the Union data set (UNION) produced by the Supernova
Cosmology Project~\cite{Kowalski:2008ez}; however, to check the consistency of
our results, we also used the recently released Constitution dataset
(Constitution)~\cite{Hicken2009} which, with 397 Type Ia supernovae, is the
largest sample to date. We also used the latest SDSS release (DR7) BAO distance
scale~\cite{Reid2009,Percival2009}. Weak lensing (WL) data are taken from
CFHTLS~\cite{Fu2007} and we use the weak lensing module provided
in ~\cite{Massey,Lesgo}, with some modifications to assess the likelihood in
terms of the variance of the aperture mass (Eq.~5 of~\cite{Fu2007}) with
the full covariance matrix~\cite{Kilbinger}. The cross-correlation between CMB
and galaxy survey data is employed using the public code at~\cite{Howebsite}. We
modify it to take into account the temporal evolution
of the dark energy equation of state, since the code only considers $w$CDM
cosmologies. We refer to~\cite{Ho1,Ho2} for a description of both the
methodology and the datasets used. Finally, we use the recent value of the Hubble constant from the SHOES
(Supernovae and $H_0$ for the Equation of State)
program,
$H_{0}=74.2\pm3.6$~km~s$^{-1}$~Mpc$^{-1}(1\sigma)$~\cite{Reiss2009}. We also incorporate baryon density
information from Big Bang Nucleosynthesis $\Omega_{b}h^{2}=0.022\pm0.002$
($1\sigma$)~\cite{Burles}, as well as a top-hat prior on the age of the
Universe, $10\mbox{ Gyr}<t_0< 20\mbox{ Gyr}$.

\begin{figure} [h]
\begin{minipage}{.5\textwidth} \centering
\epsfig{figure=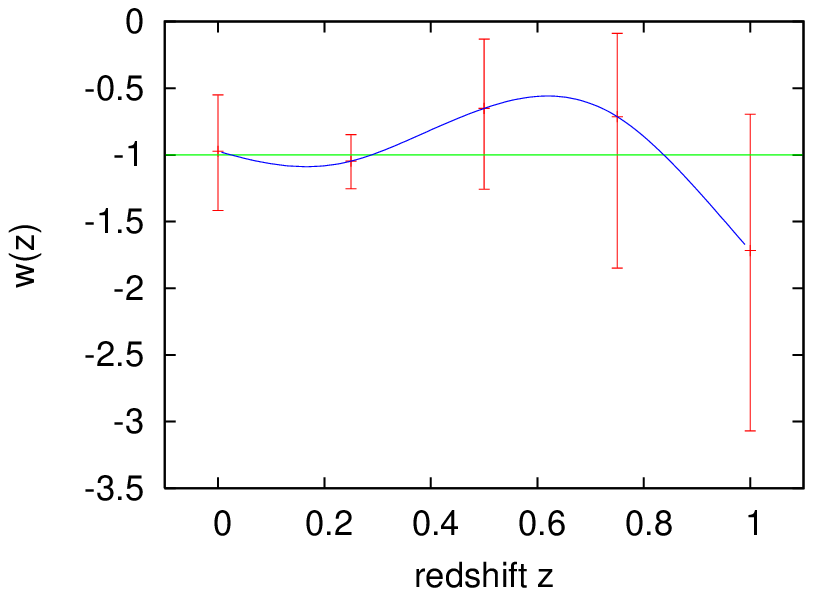,width = \textwidth}
\end{minipage} \hfill \begin{minipage}{.5\textwidth}\centering
\epsfig{figure=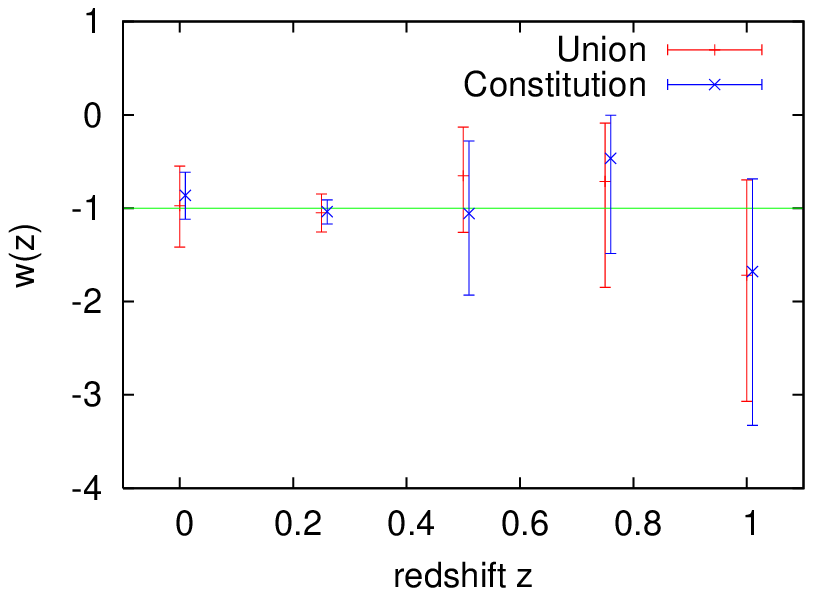, width = \textwidth}
\end{minipage}
\caption{Left: uncorrelated constraints on the dark equation of state parameters 
using WMAP+UNION+BAO. Right: comparison between
WMAP+UNION+BAO and WMAP+Constitution+BAO; the points for the
Constitution dataset have been slightly shifted to facilitate comparison
between the two cases: we find no significant difference between UNION and
Constitution.\label{fig:color}}
\end{figure}
As we can see from Fig.~1, all values are
compatible with a cosmological constant ($w=-1$) at the $2\sigma$ level; in particular, 
there is no discrepancy between the Union and Constitution datasets. Moreover, as we can see from Fig.~2, 
 the addition of cosmological probes of cosmic clustering noticeably reduces the uncertainty in the 
determination of the dark energy parameters, especially at high redshifts.

\begin{figure} [h]
\centering
\epsfig{figure=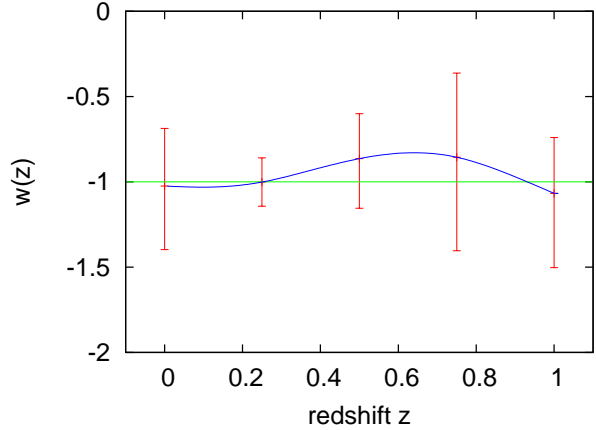,width = .53\textwidth}
\caption{Results using data from a ``global'' dataset which includes WMAP+UNION+BAO+WL+ISW+LSS; error bars are at $2\sigma$.}
\end{figure}

\begin{figure} [h]
\centering
\epsfig{figure=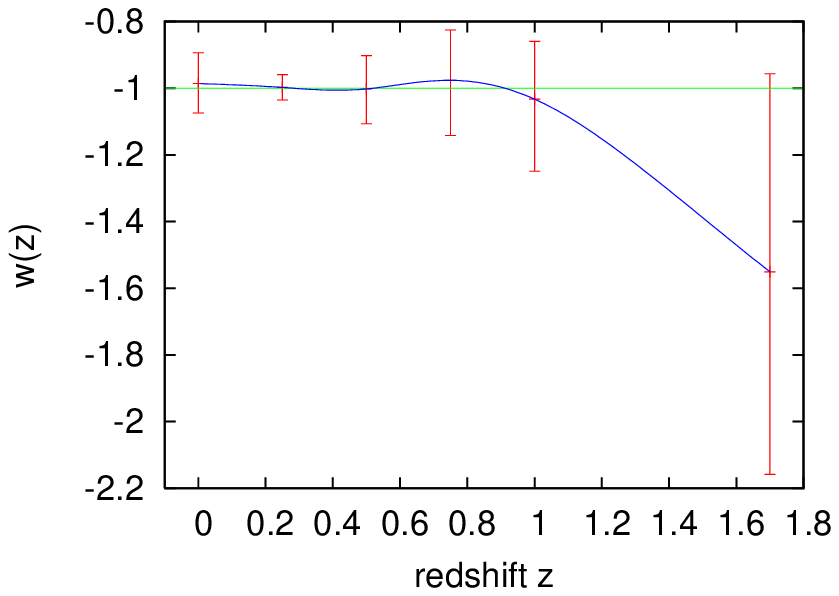,width = .53\textwidth}
\caption{Uncorrelated constraints on the dark energy equation of state parameters, for mock
datasets from Planck and JDEM; error bars are at $2\sigma$.\label{fig:redshift}}
\end{figure}
To reinforce our conclusions, we also created several mock datasets for upcoming
and future SN, BAO, and CMB experiments. The quality of future datasets allows
us to constrain the dark energy evolution beyond redshift $z=1$. We thus
consider an additional bin at $z=1.7$, with a similar constraint:
$w(z>1.7)=w(z=1.7)$. We consider a mock catalog of 2,298 SNe, with 300 SNe
uniformly distributed out to z = 0.1, as expected from ground-based low redshift
samples, and an additional 1998 SNe binned in 32 redshift bins in the range $0.1
< z < 1.7$, as expected from JDEM or similar future surveys \cite{kim:04}. The
error in the distance modulus for each SN is given by the intrinsic
error, $\sigma_{\mbox{int}}=0.1\,\mbox{mag}$. In addition, we use
a mock catalog of 13 BAO estimates, including 2 BAO estimates at z = 0.2 and
z = 0.35, with 6\% and 4.7\% uncertainties (in $D_V$), respectively, 4 BAO
constraints at $z=[0.6,0.8,1.0,1.2]$ with corresponding fiducial survey
precisions (in $D_V$) of $[1.9,1.5,1.0,0.9]\%$ (V5N5 from~\cite{Seo03}), and 7
BAO estimates with precision $[0.36,0.33,0.34,0.33,0.31,0.33, 0.32]\%$
from $z=1.05$ to $z=1.65$ in steps of 0.1~\cite{private}.\\
We simulate Planck data using a fiducial $\Lambda$CDM model, with the
best fit parameters from WMAP5, and noise properties consistent with a combination
of the Planck $100$--$143$--$217$ GHz channels of the HFI~\cite{Planck2009}, and
fitting for temperature and polarization using the full-sky likelihood function given
in~\cite{LewisPlanck}. In addition, we use the same priors on the Hubble parameter
and on the baryon density as considered above.  As can be seen
from Table~1 and Figure~3, future data will reduce the uncertainties in ${w_i}$ by a
factor of at least $2$, with the relative uncertainty below $10\%$ in all but
the last bin (at $z=1.7$).
\section{Conclusions} 
One of the main tasks for present and future dark energy surveys is to determine whether
or not the dark energy density is evolving with time.\\
We have performed a global analysis of the latest cosmological datasets and have constrained the
dark energy equation of state using a very flexible and almost model independent
parameterization. We determine the equation of state $w(z)$ in five independent
redshift bins, incorporating the effects of dark energy perturbations. We find
no evidence for a temporal evolution of dark energy---{\em the data is completely consistent
with a cosmological constant}. This agrees with most previous
results, but significantly improves the overall
constraints. We show that future 
experiments, such as Planck or JDEM, will be able to reduce the
uncertainty on $w(z)$ to less than $10\%$ in multiple redshift bins,
thereby mapping any temporal evolution of dark energy with
high precision. With this data it will be possible to measure
the temporal derivative of the equation of state parameters, ${dw}/{dz}$,
useful in discriminating between two broad classes of ``thawing'' and
``freezing'' models~\cite{Caldwell2009}.

\end{document}